\documentclass[prd,
 notitlepage,
nofootinbib,
amsfonts,
amssymb,
amsmath,
reprint,
aps,
 ]{revtex4-2}
\usepackage{color}
\usepackage{graphicx}
\usepackage{cancel}
\usepackage{mathrsfs}
\usepackage{xcolor}
\definecolor{bluc}{cmyk}{1,0.9,0,0}
\definecolor{rossoCP3}{cmyk}{0,.88,.77,.40}
\definecolor{rosso}{cmyk}{0,1,1,0.4}
\definecolor{rossos}{cmyk}{0,1,1,0.55}
\definecolor{rossoc}{cmyk}{0,1,1,0.2}
\definecolor{verdes}{cmyk}{0.92,0,0.59,0.4}
\usepackage[normalem]{ulem}
\usepackage{hyperref}
\usepackage[dvipsnames]{xcolor}
\hypersetup{colorlinks=true, bookmarksopen, bookmarksnumbered, citecolor=RubineRed, linkcolor=ForestGreen, pdfstartview=FitH, urlcolor=black}
\usepackage[utf8x]{inputenc}
\usepackage[T1]{fontenc}
\usepackage{subdepth}

\usepackage{multirow}

\newcommand\rst{\bgroup\markoverwith{\textcolor{red}{\rule[0.5ex]{1pt}{1.4pt}}}\ULon}

\begin{document}
\title{Static Black Holes and Iyer--Wald Entropy in $f(\mathcal{R},\mathcal{K})$ Gravity}
\author{I. D\'iaz-Salda\~na}
\email{ isaacdiaz@fisica.uaz.edu.mx}
\author{J. L\'opez-Dom\'inguez}
\email{ jlopez@fisica.uaz.edu.mx}
\author{Wilfredo Yupanqui}
\email{w.yunpanqui@fisica.uaz.edu.mx }
\author{Javier Chagoya}
\email{ javier.chagoya@fisica.uaz.edu.mx}
\affiliation{Unidad Acad\'emica de F\'isica, Universidad Aut\'onoma de Zacatecas,\\ Calzada Solidaridad esquina con Paseo a la Bufa S/N C.P. 98060, Zacatecas, M\'exico.}
\author{M. Sabido}
\email{ msabido@fisica.ugto.mx}
\affiliation{Departamento de F\'{\i}sica de la Universidad de Guanajuato,
 A.P. E-143, C.P. 37150, Le\'on, Guanajuato, M\'exico.}
 \affiliation{Department of Theoretical Physics, University of the Basque Country UPV/EHU,\\ P.O BOX 644, Bilbao, Spain.}
\date{\today}

\begin{abstract}
We investigate static, spherically symmetric black hole solutions in a class of {nonpolynomial}  gravitational theories described by a {Lagrangian $f=-X^{1-n}$, where $X = \mathcal{R} + \sqrt{6\mathcal{K} - \mathcal{R}^2}$ is a function of the Ricci and Kretschmann scalars.} 
The modified Einstein equations contain {nonpolynomial} curvature terms. We develop a perturbative scheme to look for a black hole solution that deviates parametrically from the Schwarzschild geometry. We obtain first-order corrections to the metric by solving the resulting coupled differential equations. The corresponding Iyer--Wald entropy is then evaluated consistently to first order in the perturbative coupling. We show that the {nonpolynomial} contribution gives rise to a power-law correction to the Bekenstein--Hawking entropy.
\end{abstract}
\maketitle


\section{Introduction}
\label{intro}
The formulation of General Relativity (GR) is one of the most outstanding achievements in theoretical physics. Nevertheless, from a theoretical perspective, it is not entirely satisfactory. For instance, the notion that every form of matter and energy gravitates is not compatible with the vacuum energy predicted by quantum field theory. {Moreover, although perturbative GR is non renormalizable when treated as a fundamental quantum field theory,  it remains meaningful as an effective theory.} On cosmological scales, the successful application of GR also relies on the introduction of a dark sector, consisting of dark matter and dark energy, to account for galaxy dynamics and the observed accelerated expansion of the Universe.  However, as observational data have become more precise,  discrepancies with the standard cosmological model are becoming relevant. In view of these circumstances, the search and study of theories of modified gravity continues to be an active area of research. 
Although there are many proposals for modified gravity, many of them share a similar philosophy:
%
 gravity is considered as a fundamental interaction, and the gravitational action is constructed following certain principles--e.g., field content, symmetries, properties of the field equations. Some examples include $f(R)$~\cite{Sotiriou:2008rp,odintsov1}, massive gravity~\cite{deRham:2014zqa}, Horndeski~\cite{Horndeski:1974wa}, etc. These theories typically address issues regarding the dark sector needed in GR, but do not have a profound impact on the problem of quantizing gravity unless they are seen as effective theories~\cite{Weinberg:2008hq,Eichhorn:2022ngh,Babichev:2025dpb}.
An alternative viewpoint is to consider gravity not as a fundamental interaction but as an emergent phenomenon. In a pioneering work, Jacobson derived Einstein's equations  from thermodynamic principles and an entropy proportional to the area~\cite{Jac}. 
More recently,  Verlinde proposed that Newtonian gravity is an entropic force (in the sense of the emergent forces present in the study of polymers) that can be derived from the entropy-area relation~\cite{Ver}. Moreover, one can derive the Friedmann equations for a Friedmann–Robertson–Walker (FRW) universe with arbitrary spatial curvature by applying the Clausius relation to the apparent horizon of the FRW universe~\cite{Cai:2005ra,Cai:2008ys}. The approach has been successful not only for GR, but also for other gravity theories~\cite{Cai:2006pa,Cai:2008ys,Chagoya:2023hjw,Chagoya:2024tqv}. 
In these approaches, the Bekenstein--Hawking (BH) entropy plays a central role. Starting from a prescribed entropy-area relation, one can derive modified Friedmann equations through the Clausius relation. A less explored question is whether the process can be reversed: given a gravitational theory and its cosmological dynamics, can one determine the entropy associated with the apparent horizon? 
More recently, the entropy has been derived directly from the modified Friedmann equations for theories of gravity whose Lagrangians contain {nonpolynomial} curvature terms~\cite{Nojiri:2023wzz,Dehyadegari:2026drv}. In particular, in~\cite{Dehyadegari:2026drv} the authors proposed a modified theory of gravity whose Lagrangian consists of the Einstein--Hilbert term supplemented by an arbitrary function $f(\mathcal{R},\mathcal{K})$ of the Ricci and Kretschmann scalars. {Such a choice for the functional dependence of $f$ is motivated by the fact that {nonpolynomial} invariants naturally arise as effective corrections to GR at high energies (see, e.g.~\cite{Alvarez-Gaume:2015rwa} and references therein). Although not essential if gravity is considered an emerging phenomenon, it is worth recalling that quadratic gravity also exhibits improved ultraviolet behavior compared to GR~\cite{Salvio:2018crh}.} This model belongs to the broad class of higher-derivative theories of gravity, whose black hole thermodynamics has been extensively investigated using several approaches, including the Iyer--Wald formalism~\cite{Dong2014,Camps2014,Miao2015,Bhattacharyya2021,Per_Kraus_2005,Charmousis2009,PhysRevLett.114.171601,Tim_Clunan_2004}.
For different choices of the {nonpolynomial} function $f(\mathcal{R},\mathcal{K})$, the authors of~\cite{Dehyadegari:2026drv} derived the corresponding modified Friedmann equations and, through the gravity--thermodynamics correspondence, identified the entropy associated with the apparent horizon of a FRW universe. In this construction, the entropy is not postulated a priori but instead emerges from the cosmological dynamics of the underlying gravitational theory. This procedure is complementary to the thermodynamic derivation of the Friedmann equations developed in~\cite{Cai:2005ra}, where one starts from a prescribed entropy-area relation and, together with the continuity equation, recovers the cosmological field equations. In particular, adopting the standard Bekenstein--Hawking entropy reproduces the Friedmann equations of General Relativity.

The existence of a well-defined entropy associated with the apparent horizon naturally raises the following question: does the same gravitational theory admit black hole solutions whose entropy is consistent with that thermodynamic description? Addressing this question constitutes the main motivation of the present work. To this end, we consider the particular $f(\mathcal{R},\mathcal{K})$ model introduced in~\cite{Dehyadegari:2026drv}, construct static, spherically symmetric vacuum black hole solutions within a first-order perturbative framework, compute the corresponding entropy using the Iyer--Wald Noether-charge formalism~\cite{Wald:1993nt,Iyer:1994ys,Iyer:1995kg}, and compare the resulting entropy-area relation with that previously obtained from the thermodynamics of the apparent horizon. In this way, we examine the consistency between these two independent thermodynamic descriptions of the same gravitational theory.

This paper is organized as follows. In Section~\ref{sec:action_equations} we introduce the modified theory of gravity and derive the corresponding field equations. Section~\ref{sec:perturbative_framework} is devoted to the construction of first-order perturbative black hole solutions. In Section~\ref{sec:wald_entropy} we compute the associated Iyer--Wald entropy, while Section~\ref{conclusions} contains our concluding remarks and outlook.
\section{Action and Field Equations}
\label{sec:action_equations}
{The relationship between gravitation and horizon thermodynamics has motivated the construction of modified theories of gravity from prescribed entropy-area relations. In a recent work~\cite{Dehyadegari:2026drv}, a class of $f(\mathcal{R},\mathcal{K})$ theories was proposed and its cosmological implications were investigated. These theories are described by the action
\begin{equation}
    \mathcal{S}= \frac{1}{16\pi G}\int d^4x \sqrt{-g} \left[ {\mathcal{R}} + \frac{\lambda}{ \ell^{2n} }f(\mathcal{R}, \mathcal{K}) \right], \label{eq:action}
\end{equation}
where $\mathcal{R}=g^{\mu\nu}\mathcal{R}_{\mu\nu}$ and $\mathcal{K} = \mathcal{R}_{\alpha\beta\gamma\delta}\mathcal{R}^{\alpha\beta\gamma\delta}$ are the Ricci and Kretschmann scalars, respectively, {$\ell$ is a characteristic length scale that controls the strength of the nonpolynomial curvature corrections, and $\lambda$ is a dimensionless coupling constant necessary to make a perturbative analysis valid in the case $n=0$. For other values of $n$, both $\lambda$ and $\ell$ can be used  equivalently to keep the contributions of $f$ in a perturbative regime.} By deriving the corresponding Friedmann equations for different choices of the nonpolynomial curvature function $f(\mathcal{R},\mathcal{K})$, the authors identified the entropy associated with the apparent horizon of a FRW universe through the gravity-thermodynamics correspondence. Among the different choices of the nonpolynomial curvature function considered in~\cite{Dehyadegari:2026drv}, we focus on
\begin{equation}
    f(\mathcal{R}, \mathcal{K}) = - X^{1-n}, \quad \quad X = \mathcal{R} + \sqrt{6\mathcal{K} - \mathcal{R}^2}, \label{eq:function}
\end{equation}
where $n$ is a dimensionless parameter controlling the order of the nonpolynomial curvature corrections.\footnote{For $n=0$, $f(\mathcal R, \mathcal K)$ is of the same order as the Einstein-Hilbert term, so it can be seen as a small correction only if $|\lambda|\ll1$.} The particular combination of the Ricci and Kretschmann scalars used in $X$ is such that the modified Friedmann equations do not contain derivatives of order higher than 2. This model was shown to generate a generalized entropy-area relation for the apparent horizon in the cosmological setting\footnote{{The entropy of the apparent horizon found within this theory, when employed in Verlinde's approach to gravitation, gives the MOND theory of gravity~\cite{Modesto:2010rm,Diaz-Saldana:2018ywm}}}.
In this work, rather than focusing on cosmological solutions, our goal is to determine static, spherically symmetric vacuum solutions and study their thermodynamic properties using the Iyer--Wald formalism. }
Let us proceed by obtaining the field equations for the action Eq.~\eqref{eq:action} with the choice Eq.~\eqref{eq:function}. Variation of the action 
with respect to $g^{\mu\nu}$ yields the generalized field equations
\begin{equation}
    G_{\mu\nu} + \frac{\lambda}{ \ell^{2n} }\, \Sigma_{\mu\nu}=0, \label{eq:general_field_eqs}
\end{equation}
where $G_{\mu\nu}$ is the Einstein tensor, and the effective source tensor $\Sigma_{\mu\nu}$ is given by
{
\begin{align}\label{eq:sigma tensor}  
 \Sigma_{\mu\nu}=&f_{\mathcal{R}} \mathcal{R}_{\mu\nu} - \frac{1}{2}g_{\mu\nu}f + \left( g_{\mu\nu} \square - \nabla_\mu \nabla_\nu \right) f_{\mathcal{R}}\\ \nonumber
 +& 2f_{\mathcal K}R_{\mu\alpha\beta\gamma}
R_\nu{}^{\alpha\beta\gamma}
+4\nabla_\alpha\nabla_\beta
\left(f_{\mathcal K}R_\mu{}^\alpha{}_\nu{}^\beta\right),
\end{align}}
with $\square\equiv\nabla_{\mu}\nabla^{\mu}$. The derivatives $f_{\mathcal{R}} \equiv \partial f / \partial \mathcal{R}$ and $f_{\mathcal{K}} \equiv \partial f / \partial \mathcal{K}$ encompass all the higher-order curvature contributions derived from the functional variation of $f(\mathcal{R}, \mathcal{K})$.

\section{First-Order Perturbative Framework and Black Hole Solutions}\label{sec:solutions}
\label{sec:perturbative_framework}
Due to the highly non-linear nature of the generalized field equations
, finding exact vacuum solutions  becomes analytically  intractable. Instead, we exploit the fact that the {nonpolynomial} sector is controlled by the coupling constant $\lambda${--in the case $n=0$--and by the characteristic scale $\ell$ when $n\neq0$. The specific way in which $\ell$ suppresses the deviations from the vacuum solutions of GR depends on the sign of $n$, as will be appreciated in the solutions that we derive below. With this observation in mind, we assume that the corrections due to the non-polynomial terms can be treated as perturbations of a GR background.} This strategy has been successfully employed in several nonpolynomial curvature theories~\cite{Bueno:2016ypa,Yunes:2011we,Bueno:2016xff}. Accordingly, we expand the spacetime metric as follows
\begin{equation}\label{eq:metric perturbation}
    g_{\mu\nu}
    =
    g_{\mu\nu}^{(0)}
    +
    \lambda\, h_{\mu\nu}
    +
    \mathcal O(\lambda^2),
\end{equation}
where $g_{\mu\nu}^{(0)}$ denotes the background geometry and $h_{\mu\nu}$ is the first-order metric perturbation. The Einstein tensor and the effective source tensor are therefore expanded as
\begin{align}
    &G_{\mu\nu} =G_{\mu\nu}^{(0)}+ \lambda\,\delta G_{\mu\nu}
    +\mathcal O(\lambda^2), \label{eq:Einstein_expansion}\\ 
     &\Sigma_{\mu\nu} =\Sigma_{\mu\nu}^{(0)}+ \lambda\,\delta \Sigma _{\mu\nu}
    +\mathcal O(\lambda^2).\nonumber
\end{align}
From now on, the subscript $(0)$ denotes evaluation on the 
background. Since our goal is to investigate static, spherically symmetric vacuum solutions, we choose the Schwarzschild spacetime as the background geometry, for which the Ricci scalar and Ricci tensor identically vanish, $\mathcal{R}^{(0)} = 0$, $\mathcal{R}_{\mu\nu}^{(0)} = 0$. Consequently, the Einstein tensor also vanishes, 
$G^{(0)}_{\mu\nu}=0$. Furthermore, for the Schwarzschild geometry, we have $ X_0 = \sqrt{6\mathcal{K}^{(0)}}$, where $\mathcal{K}^{(0)}=48M^2/r^6$ is the corresponding Kretschmann scalar. Therefore, the derivatives of the nonpolynomial curvature function evaluated on the background reduce to
\begin{align}
    f_{\mathcal{R}}^{(0)} = -(1-n) X_0^{-n}, \quad
    f_{\mathcal{K}}^{(0)} = -3(1-n) X_0^{-(n+1)}. \label{eq:fR_scaling}
\end{align}
Substituting Eq.~\eqref{eq:Einstein_expansion} and keeping only terms up to $\mathcal O(\lambda)$, the field equations reduce to
\begin{equation}
    \delta G_{\mu\nu}=-\frac{1}{ \ell^{2n} }\Sigma_{\mu\nu}^{(0)}.\label{eq:first_order_eqs}
\end{equation}
We now specialize to static, spherically symmetric solutions of Eq.~\eqref{eq:first_order_eqs}. To this end, we consider the metric ansatz
\begin{align}
    ds^2 = &- P(r) dt^2 +  Q(r) dr^2 + r^2 d\Omega^2, \label{eq:metric}
\end{align}
with
\begin{align}
    P(r) =   \mathscr{F}_0(r) + \lambda\, A(r), \quad
    Q(r)=\left[ \mathscr{F}_0(r) + \lambda\, B(r)\right]^{-1},
\end{align}
where $\mathscr{F}_0(r)=1-2M/r$ and $A(r)$, $B(r)$ are unknown functions. 
Substituting the metric ansatz into 
 Eq.~\eqref{eq:first_order_eqs}, written in mixed form $\delta G^{\mu}_{\ \ \nu}=- (1/\ell^{2n})\Sigma^{(0)\mu }_{\quad \ \nu}$, we obtain the first-order differential equations
\begin{align}
   & \frac{1}{r^2} \frac{d}{dr}\big[r B(r)\big] = -\ell^{-2n}\Sigma^{(0)t }_{\quad \ t}, \label{eq:Gtt} \\
    &\frac{r B(r)-2M A(r)}{r^{2}(r-2M)} + \frac{1}{r} A'(r) = -\ell^{-2n}\Sigma^{(0)r }_{\quad \ r}.\nonumber
\end{align}
The quantities $\Sigma^{(0)t }_{\quad \ t}$ and $\Sigma^{(0)r }_{\quad \ r}$ can be straightforwardly calculated from Eq.~\eqref{eq:sigma tensor}. They can be written in the compact form
\begin{align}
   &\Sigma^{(0)t }_{\quad \ t}(r)=\kappa_{n}M^{-n}r^{3(n-1)}(\alpha_{n}M-\beta_{n}r),\\
   & \Sigma^{(0)r }_{\quad \ r}(r)=\kappa_{n}M^{-n}r^{3(n-1)}(\psi_{n}M+\sigma_{n}r),\nonumber
\end{align}
where we have introduced the coefficients 
{\begin{align}
   & \kappa_{n}=2^{-\frac{5n}{2}} \, 3^{1-n}\\
   &\alpha_n = 2\sqrt{2} - n(n-1)\left[1+7\sqrt{2}+6(1+\sqrt{2})n\right],\\
   &\beta_n = -(n-1)(3n+1)\left[\sqrt{2}+(1+\sqrt{2})n\right],\\
   & \psi_n = -2\sqrt{2} + n(3+\sqrt{2}+3(\sqrt{2}-1)n),\\
   &\sigma_n = -(n-1)\left[\sqrt{2}+(\sqrt{2}-2)n\right].
\end{align}}
{Note that the covariant divergence of the Riemann tensor in Eq.~\eqref{eq:sigma tensor} vanishes for the Schwarzschild background, as can be shown using the Bianchi identities and the fact that $R^{(0)}_{\mu\nu}=0$. Thus, on the background, 
$[\nabla^\alpha \nabla^\beta (f_{\mathcal{K}} \mathcal{R}_{\mu\alpha\nu\beta})]^{(0)} = [(\nabla^\alpha \nabla^\beta f_{\mathcal{K}}) \mathcal{R}_{\mu\alpha\nu\beta}]^{(0)}$.}

The system formed by Eq.~\eqref{eq:Gtt} can be solved analytically. For $n\neq0$, the resulting metric perturbations are
\begin{align}
&A(r) = - \frac{\kappa_n  M^{-n} }{3n \ell^{2n}}r^{3n-1} \left( \psi_n M + \omega_{n} r \right) + \frac{c_1}{r} + c_2 \mathscr{F}_0(r),\label{eq:solA}\\
&B(r) = - \frac{\kappa_n  M^{-n}}{3n \ell^{2n}} r^{3n-1} \left( {\alpha_n} M - \frac{3n}{3n+1} \beta_n r \right) + \frac{c_1}{r}, \label{eq:solB}
\end{align}
where $\omega_{n}= \sigma_n + \frac{\beta_n}{3n+1}$
and $c_1, c_2$ are integration constants. {The constant $c_2$ can be removed by a redefinition of the time coordinate--the same argument used to remove one of the integration constants that appear when finding the solution $\mathscr F_0$. On the other hand, since $P(2M) = Q(2M)$, a bifurcate Killing horizon can be established at $r=2M$ by choosing
\begin{align}
  c_1= \frac{2^{n/2} 3^{-n} \left[\left(\sqrt{2}+1\right) (1-n) n+2 \sqrt{2}\right] \ell ^{-2 n} M^{2 n+1} }{n}.\label{eq:c1nneq0}
\end{align}
This horizon is non-degenerate, with surface gravity given by
\begin{equation}
    \kappa = \frac{1}{4 M} - \frac{\lambda   \left[\left(\sqrt{2}+2\right) n^2-\left(\sqrt{2}-6\right) n+4\right] M^{2 n-1} }{3^n 2^{(7-n)/2}n \ell^{2n}} + \mathcal O(\lambda^2),
\end{equation}
where we have presented the result as a series in $\lambda$ only for simplicity. Therefore, the application of Wald's formalism to compute the entropy is justified. Notice that for $n=1$ the solution reduces to $A(r)=B(r)=-r^2/6\ell^2 + 4 M^3/(3 r \ell^2)$, thus obtaining the Schwarzschild--(A)dS metric~\cite{Kottler:1918cxc} after identifying an effective cosmological constant $\Lambda_{eff} =\lambda/2\ell^2$ and a mass parameter $\tilde M=M-\frac{2 \lambda  M^3}{3 \ell ^2}$. This result is expected since, as can be seen from the action in Eq.~\eqref{eq:action}, for $n=1$ the modification becomes $-\lambda$. This also clarifies the role of $c_1$ as a constant used to keep a simple expression for the horizon radius at the expense of redefining the mass.  Note that near the horizon $\lambda c_1/r \sim \lambda (M/\ell)^{2n}$, which is of the same 
order as the remaining terms in $\lambda A(r)$ and $\lambda B(r)$. Therefore, this choice of $c_1$ is consistent with the 
observation that, for $n\neq0$, the perturbative expansion in our region of interest can be is controlled by $|\lambda|(M/\ell)^{2n}$.}\footnote{{This can also be found by imposing the reality of $\sqrt{6\mathcal{K}-\mathcal{R}^2}$ to first order in $\lambda$ near the horizon, since $6\mathcal{K} - \mathcal{R}^2 \approx 6\mathcal{K}^{(0)} + 6\lambda \delta \mathcal{K}$ and therefore $| \lambda \delta \mathcal{K}(2M) | \ll \mathcal{K}^{(0)}(2M)$. For our metric,  Eq.~\eqref{eq:metric},  $| \lambda\delta \mathcal{K}(2M) | \sim  | \lambda |\frac{3}{4M^4}\left(\frac{M}{\ell}\right)^{2n}$ and it follows  $|\lambda| \left(\frac{M}{\ell}\right)^{2n} \ll 1$.}}, 
Moreover, a careful analysis of these closed-form solutions reveals that the leading radial terms scale asymptotically as $\mathcal{O}(r^{3n})$, except in the case $n=1$ where it scales as $\mathcal{O}(r^{3n-1})$. Therefore, for the spacetime to remain asymptotically flat, the scaling parameter must satisfy {$n <- 1/3$}\footnote{{The same conclusion is obtained by requiring $A$ and $B$ to be smaller than the background term $M/r$.}}. Under this physical regime, the modifications to the metric naturally vanish as $r \to \infty$, safely recovering the Minkowski limit while 
{perturbatively} altering the near-horizon regime where the modifications are relevant. {On the other hand, when $n>-1/3$ the functions $A(r), B(r)$ grow asymptotically and the perturbative regime breaks down for sufficiently large $r$. However, the analysis of the entropy presented in the next section does not rely on the asymptotic properties of the spacetime}.
For the particular case $n=0$, the metric perturbations are given by
\begin{align}
    A(r) &= \frac{c_1+6\sqrt{2}M  - 2Mc_2 + rc_2 + 6\sqrt{2}(M - r)\log\left(\frac{r}{r_0}\right)}{r}, \\
    B(r) &= \frac{c_1 + 3\sqrt{2}\left[r - 2M\log\left(\frac{r}{r_0}\right)\right]}{r},
\end{align}
where $c_1$, $c_2$ are integration constants, {and $r_0$ is an arbitrary length scale which can be set as $r_0=2M$. Imposing that the constant part of $A(r)$ vanishes for large $r$, we set $c_2=0$.} On the other hand, {to first order, the corrected horizon is given by} 
{\begin{equation}
r_H=2M-\lambda\left[ 6\sqrt2\,M +c_1\right],\label{eq:rh}
\end{equation}
alternatively, we can use $c_1$ to maintain the horizon at $r=2M$. Indeed, the corrections $A_{n=0}$ and $B_{n=0}$ vanish at $r=2M$ for
\begin{align}
c_1 =- 6\sqrt2\,M .\label{eq:c1n0}
\end{align}}
{Thus, for this choice of $c_1$, the trapping horizon ($g^{rr}=0$, the one we associate with the event horizon) coincides with the Killing horizon ($g_{tt}=0$) at $r=2M$, providing a bifurcate Killing horizon where we will apply Wald's entropy formalism in the next section.}
Although conditions~\eqref{eq:c1nneq0} and~\eqref{eq:c1n0} guarantees that there is a horizon at $r=2M$, the spacetime metric may have more horizons, and whether $r=2M$ corresponds to the exterior black hole horizon (where $g^{rr}$ and $g_{tt}$ change from positive to negative as they enter the horizon) depends on the values of $\lambda$. 

The analysis above highlights a vast diversity of behaviors of the perturbative corrections to the Schwarzschild black hole. A detailed analysis of these corrections is left for future work since, as we show below, the black hole entropy computed in the Iyer-Wald formalism is not sensitive to the details of the corrections; for the purpose of applying the Iyer-Wald formalism, our main conclusion so far is that the theory defined by Eq.~\eqref{eq:action} admits black hole solutions with trapping and Killing horizons that coincide for appropriate values of the integration constant. 

\section{Black Hole Entropy from the Iyer--Wald Formalism}
\label{sec:wald_entropy}

Having established the perturbative solutions for the modified geometry, we now turn our attention to the thermodynamic implications of the higher-order curvature corrections. To compute the entropy of the modified black hole, we employ the Iyer--Wald formalism. The Iyer--Wald entropy is proportional to the Noether charge associated with the diffeomorphism invariance of the theory, evaluated strictly on the bifurcation surface of the Killing horizon $\mathcal{H}$. Since the Noether charge is evaluated locally on the bifurcation surface, the resulting entropy is a quasi-local quantity that depends only on the geometry of the Killing horizon and not on the asymptotic structure of the spacetime~\cite{Jacobson:1993vj,Wald:1999vt}.\footnote{{The asymptotic form of the metric would be relevant if we wanted to construct the first law following Wald's procedure, since, for instance, the definition of the total mass depends on the asymptotic properties of the spacetime. Here, we are only concerned with the entropy as a Noether charge. {A rigorous derivation of the first law requires a phase-space analysis within the covariant phase-space formalism of Wald, including an assessment of the relevant fluxes at the horizon and at infinity. We leave this analysis for future work.}}} 
{For a general diffeomorphism-invariant theory described by a covariant $D$-form Lagrangian $\boldsymbol{L}$, the Iyer--Wald entropy is given by
\begin{equation}
    S_W = -2\pi \oint_{\mathcal{H}} \frac{\partial L}{\partial \mathcal{R}_{\mu\nu\alpha\beta}} {\epsilon}_{\mu\nu} {\epsilon}_{\alpha\beta} \sqrt{h}\,d^{D-2}x.
    \label{eq:wald_formula}
\end{equation}
Here, $L$ denotes the scalar Lagrangian associated with the covariant $D$-form $\boldsymbol{L}=L\, \boldsymbol{\mathrm{vol}}_D$, $h$ is the determinant of the induced metric on the bifurcation surface, ${\epsilon}_{\mu\nu}=- {\epsilon}_{\nu\mu}$ is the binormal tensor to the bifurcation surface, normalized such that ${\epsilon}_{\mu\nu}{\epsilon}^{\mu\nu} = -2$ and the functional derivative is taken considering the metric and the Riemann tensors as independent fields. For the theory defined by the action in Eq.~\eqref{eq:action}, the derivative entering Eq.~\eqref{eq:wald_formula} is given by
\begin{align}
    \frac{\partial L}{\partial \mathcal{R}_{\mu\nu\alpha\beta}} &= \frac{1}{16\pi G} \left[  \mathcal{C}^{\mu\nu\alpha\beta}
    + \frac{\lambda}{\ell^{2n}} \left (f_{\mathcal{R}}\,\mathcal{C}^{\mu\nu\alpha\beta} + 2 f_{\mathcal{K}} \mathcal{R}^{\mu\nu\alpha\beta} \right)\right],
    \label{eq:lagrangian_derivative}
\end{align}
with $\mathcal{C}^{\mu\nu\alpha\beta}\equiv \frac{1}{2}(g^{\mu\alpha}g^{\nu\beta} - g^{\mu\beta}g^{\nu\alpha})$.}
{Contracting with $\epsilon_{\mu\nu}\epsilon_{\alpha\beta}$ and using the normalization condition $\epsilon_{\mu\nu}\epsilon^{\mu\nu}=-2$, the purely Einstein--Hilbert contribution reduces to
\begin{equation}
S_{GR}
=
-2\pi
\left(
\frac{1}{16\pi G}
\right)
(-2)
\oint_{\mathcal H}
\sqrt{h}\,
d^2x
=
\frac{\mathcal A_H}{4G},
\label{eq}
\end{equation}
obtaining the standard BH entropy, as expected.}
The curvature invariants evaluated on the perturbed horizon differ from their Schwarzschild values only by terms of order $\mathcal{O}(\lambda)$. Since the nonpolynomial curvature contribution to the Wald entropy, namely $S_{NP}$, is itself proportional to $\lambda$, corrections induced by the metric perturbation contribute only at order $\mathcal{O}(\lambda^2)$. Therefore, throughout the present first-order analysis, the geometric quantities entering $S_{NP}$ can consistently be evaluated on the unperturbed Schwarzschild background. In particular, $\mathcal{A}_H^{(0)}=16\pi M^2$. Using the covariant expression for the binormal contraction, the nonpolynomial curvature contribution can be written as
\begin{equation}
\begin{aligned}
S_{NP}
=
-\frac{\lambda}{4G\ell^{2n}}
&\oint_{\mathcal H}\sqrt{h}\,d^2x
\bigg[
-f_{\mathcal R}^{(0)}\\
&+
f_{\mathcal K}^{(0)}
\left(
\epsilon_{\mu\nu}\epsilon_{\alpha\beta}
\mathcal R^{\mu\nu\alpha\beta}
\right)^{(0)}
\bigg]_{ r= r^{(0)}_{H}}.
\label{eq:Swald}
\end{aligned}
\end{equation}
Since the Schwarzschild background is spherically symmetric, the quantities $f_{\mathcal R}^{(0)}$, $f_{\mathcal K}^{(0)}$, and $\left(\epsilon_{\mu\nu}\epsilon_{\alpha\beta}\mathcal R^{\mu\nu\alpha\beta}\right)^{(0)}$ are constant over the horizon two-sphere. Consequently, they can be factored out of the surface integral, leaving only the horizon area. Thus,
\begin{equation}
    S_{NP}
    =
    -\frac{\lambda}{4G\ell^{2n}}
    \left[
    -f_{\mathcal R}^{(0)}
    +
    f_{\mathcal K}^{(0)}
    \left(
    \epsilon_{\mu\nu}\epsilon_{\alpha\beta}
    \mathcal R^{\mu\nu\alpha\beta}
    \right)^{(0)}
    \right]_{ r=r^{(0)}_{H}}
    \mathcal A_{H}^{(0)}.
    \label{eq:entropy_mod_factored}
\end{equation}
For the static, spherically symmetric metric~\eqref{eq:metric}, the normalization condition $\epsilon_{\mu\nu}\epsilon^{\mu\nu}=-2$ yields
$\epsilon_{tr}=-\epsilon_{rt}=\pm\sqrt{P(r)Q(r)}$. Choosing the positive orientation, we take $\epsilon_{tr}=\sqrt{P(r)Q(r)}$. Thus, on the unperturbed Schwarzschild background we have $P_0(r)Q_0(r)=1$, so that $\epsilon_{tr}^{(0)}=-\epsilon_{rt}^{(0)}=1$. At the horizon, this result is understood through the appropriate limiting prescription in the static coordinates. Therefore, using the Schwarzschild background we find
\begin{equation}
\left(
\epsilon_{\mu\nu}\epsilon_{\alpha\beta}
\mathcal R^{\mu\nu\alpha\beta}
\right)^{(0)}_{ r= r^{(0)}_{H}}=-\frac{16\pi}{\mathcal A_H^{(0)}}.
\end{equation}
Substituting this, along with 
\begin{align}
&\left. f_{\mathcal R}^{(0)} \right|_{r=r_H^{(0)}}
= -(1-n)\left(
\frac{\mathcal A_H^{(0)}}{24\sqrt{2}\,\pi}
\right)^n,
\\
&\left. f_{\mathcal K}^{(0)} \right|_{r=r_H^{(0)}}
= -3(1-n)\left(
\frac{\mathcal A_H^{(0)}}{24\sqrt{2}\,\pi}
\right)^{n+1}
\end{align}
in Eq.~\eqref{eq:entropy_mod_factored}, we find
\begin{equation}
    S_{NP} = - \frac{\lambda}{\ell^{2n} G} {2^{-5/2-7n/2}(3\pi)^{-n}(1-n)(2+\sqrt{2})} \mathcal{A}_H^{n+1}. \label{eq:mod_entropy_exact}
\end{equation}
 Note that, to first order in the coupling parameter $\lambda$, replacing the background horizon area $\mathcal{A}_H^{(0)}$ with the full perturbed horizon area $\mathcal{A}_H$ introduces corrections only at order $\mathcal{O}(\lambda^2)$. Therefore, we express the final result for $S_{NP}$ in terms of $\mathcal{A}_H$. Putting everything together, the total Iyer--Wald entropy takes the form
\begin{equation}
    S_W = \frac{\mathcal{A}_H}{4G}+\frac{\lambda}{\ell^{2n}}
    \frac{(n-1)(1+\sqrt{2})G^n}{(6\sqrt{2}\pi)^n}\left(\frac{\mathcal{A}_H}{4G}\right)^{n+1}.
    \label{eq:final_wald_entropy}
\end{equation}
Notably, the nonpolynomial curvature contribution appears as a power of the BH entropy. {The comparison between  the stationary Wald entropy and the entropy obtained from the cosmology at the apparent-horizon \footnote{{Rewriting the entropy in \cite{Dehyadegari:2026drv} to match our convention for the horizon area, gives  $S_h =\frac{\mathcal{A}_H}{4G}+\frac{2^{1-2n}(n-1)(2n-1)\lambda G^n}{3^n(n+1)\pi^n}\left(\frac{\mathcal{A}_H}{4G}\right)^{n+1}$.}},  reveals both a striking structural correspondence as well as fundamental differences. Although both formulas share the identical power law scaling $S^{n+1}$ and both vanish at $n=1$ the coefficients are different and the limiting behaviors for some values of $n$ differ significantly. Most notably, the cosmological expression introduces a $(2n-1)$ causing it to vanish for $n=1/2$ and a factor $(n+1)$ in the denominator yielding a singularity  at $n=-1$, neither of which appear in stationary result. For $n=1/2$, the discrepancy can be traced to  the choice of  temperature used in the dynamical cosmological setting, giving a differential scaling of the dynamical volume expansion $(2n-1)r_A^{2n-2}$
under the Clausius relation.
}
Interestingly, the choice $n=1/2$ leads to an entropy correction proportional to $\mathcal{A}_{H}^{3/2}$, which was shown in~\cite{Chagoya:2023hjw,Diaz-Saldana:2018gxx} to produce a modified Friedmann equation describing a self-accelerating universe. Furthermore, in~\cite{Chagoya:2023hjw}, this cosmological model was found to be equivalent to the Dvali--Gabadadze--Porrati (DGP) model~\cite{Dvali:2000hr,Deffayet:2001pu,Deffayet:2008zza,Xia_2009}.
{For $n=-1$, the discrepancy can be traced to the temporal integration over background cosmic expansion within the Clausius relation which introduces a denominator factor $n+1$. 
The absence of the singularity $n=-1$ for the static case is  particularly interesting}. In this case, the nonpolynomial curvature contribution scales as $\mathcal{A}_H^{(0)}$, so the entropy correction reduces to a pure constant, completely independent of the horizon area. 
{Moreover, since the Wald entropy is evaluated on the unperturbed Schwarzschild background, the background scalar invariant reduces to the Kretschmann scalar. However, because the derivative with respect to the Riemann tensor does not vanish, the entropy density incorporates non-trivial boundary contributions beyond a naive Gauss–Bonnet replacement, yielding a modified horizon value and reproducing a constant entropy shift characteristic of four-dimensional Einstein--Gauss--Bonnet gravity. Crucially, this correspondence is strictly restricted to the perturbative evaluation of the entropy functional: the field equations remain non-trivial for $n=-1$, yielding non-vanishing corrections to the metric functions. Therefore, the present theory is not dynamically equivalent to Einstein gravity supplemented by a topological Gauss–Bonnet term, even though both theories share a constant contribution to the black hole entropy within the first-order perturbative regime.}

{It is also instructive to consider the neighborhood of the distinguished value $n=-1$. Writing $n=-1+\varepsilon$, with $|\varepsilon|\ll1$, and expanding the full entropy correction in Eq.~\eqref{eq:final_wald_entropy}, we obtain
\begin{equation}
\begin{aligned}
S_W={}&S_{BH}
-\frac{2\lambda(1+\sqrt{2})}{q}\\
&+\frac{2\lambda\varepsilon(1+\sqrt{2})}{q}
\left[
\frac{1}{2}-\log q- \log S_{BH}
\right]
+\mathcal{O}(\varepsilon^2),
\end{aligned}
\end{equation}
where $q=G/(6\sqrt{2}\pi\ell^2)$. Thus, the first-order correction contains both an area-independent contribution and a logarithmic term. The latter represents the area-dependent part that emerges in the expansion around $n=-1$, whereas the entropy correction at $n=-1$ itself is constant. This logarithmic dependence follows directly from the analytic structure of the entropy formula and requires no additional assumptions beyond the perturbative expansion in $\varepsilon$.}

The logarithmic correction to the entropy has been found in several modified theories of gravity, including a recent study employing the Iyer--Wald formalism in $f(\mathcal R)$ gravity~\cite{Mondal:2026mqb}, as well as in vector-tensor theories~\cite{Chagoya:2023ddb} and in a variety of quantum-gravity-inspired approaches~\cite{Carlip:2000nv,Kaul:2000kf,Obregon:2000zd,Sen:2012dw,Mukherji:2002de}, to cite only a few representative examples. Although the underlying dynamics of these theories differ substantially from those considered here, the emergence of the logarithmic term suggests that it can be understood as a distinguished mathematical limit of the more general power-law entropy correction derived in the present work.
{Finally, the case $n=0$ also deserves special attention since 
the total entropy remains proportional to the standard BH entropy. In other words, unlike the other cases, there is  no  new geometric dependence in the entropy.} {This seems to suggest that for $n=0$  the nonpolynomial curvature contribution effectively changes the gravitational coupling rather than introducing new geometrical information, however, as we learnt in the case $n=-1$, the fact that the (perturbative) entropies of two theories coincide does not necessarily imply a full equivalence between them. Indeed, the black hole solution in the $n=0$ case is not a solution of vacuum GR. }
{
As a consistency check, we verify that the total first-order correction to the Wald entropy in Eq.~\eqref{eq:final_wald_entropy} satisfies the general thermodynamic relation derived by Reall and Santos~\cite{Reall:2019sah} for arbitrary higher-derivative gravity theories. It is important to remark that for a Schwarzschild background, the application of this general Reall-Santos identity requires the perturbed spacetime to remain asymptotically flat; therefore, we restrict ourselves to the branch $n < -1/3$. In this regime, setting $c_1 = 0$ in our solutions\footnotemark~Eqs.~\eqref{eq:solA} and \eqref{eq:solB} ensures that the parameter $M$ coincides with the physical ADM mass. The total entropy correction must be decomposed as $\delta S_W = \delta S_{\mathrm{area}} + \delta S_{NP}$, where $\delta S_{\mathrm{area}} = 4\pi M \delta r_H$ accounts for the change in the horizon area (setting $G=1$), while $\delta S_{NP}$ represents the contribution of the nonpolynomial sector to the Noether charge, evaluated on the unperturbed horizon.

To first order, the shift in the horizon radius is determined by the radial metric perturbation derived in Sec.~\ref{sec:solutions}. Imposing the horizon condition gives $\delta r_H = -2M\lambda\,B(2M)$. Substituting the corrected metric function Eq.~\eqref{eq:solB} 
\begin{equation}
    \delta r_H = -\frac{\lambda M}{n}
    \left(\frac{\sqrt{2}M^2}{3\ell^2}\right)^n
    \left[(1+\sqrt{2})(n^2-n) - 2\sqrt{2}\right],
\end{equation}
then, the corresponding area contribution is
\begin{equation}
    \delta S_{\mathrm{area}}
    = -\frac{4\pi\lambda M^2}{n}
    \left(\frac{\sqrt{2}M^2}{3\ell^2}\right)^n
    \left[(1+\sqrt{2})(n^2-n) - 2\sqrt{2}\right].
\end{equation}
The nonpolynomial contribution evaluated on the unperturbed horizon is
\begin{equation}
    \delta S_{NP} = 4\pi\lambda M^2
    \left(\frac{\sqrt{2}M^2}{3\ell^2}\right)^n
    (n-1)(1+\sqrt{2}).
\end{equation}
Using $(n^2-n)/n=n-1$, the terms proportional to $(n-1)(1+\sqrt{2})$ cancel when the two contributions are added. The total first-order entropy correction therefore reduces to
\begin{equation}
    \delta S_W = \frac{8\sqrt{2}\pi}{n}\lambda M^2
    \left(\frac{\sqrt{2}M^2}{3\ell^2}\right)^n,
    \qquad n < -1/3.
\end{equation}
This agreement provides a nontrivial consistency check of our perturbative solutions against the Reall--Santos relation in the asymptotically flat regime considered here.
}

\footnotetext{{Note that in Sec.~\ref{sec:solutions}, the integration constant $c_1$ was chosen to keep the coordinate location of the horizon fixed at $r=2M$, effectively absorbing the horizon displacement into a redefinition of the mass parameter. In the present thermodynamic analysis, we instead set the homogeneous constant to zero $c_1=0$ to work at fixed ADM mass. In the asymptotically flat regime considered here, this choice ensures that $M$ coincides with the ADM mass to first order in $\lambda$. The geometric effect of the perturbation then manifests itself explicitly as a shift of the horizon radius, $\delta r_H$.}}
\section{Outlook and final remarks}\label{conclusions}
{In this work, we have investigated static, spherically symmetric black hole solutions in a novel class of $f(\mathcal{R},\mathcal{K})$ theories, where the gravitational action is constructed from a nonlinear combination of the Ricci and Kretschmann scalars. Treating the nonpolynomial curvature sector perturbatively around the Schwarzschild geometry, we derived the modified field equations and obtained first-order solutions for the metric functions. These solutions provide a fully analytical description of the leading nonpolynomial curvature corrections and reveal a remarkably rich dependence on the parameter $n$, with distinct geometric and thermodynamic behaviors emerging in different regions of the parameter space.

From the geometrical perspective, the perturbative solutions show that asymptotic flatness is preserved only for $n<0$, while the particular choice $n=1$ exactly reproduces the Schwarzschild--(anti--)de Sitter geometry, where the nonpolynomial curvature interaction effectively behaves as a cosmological constant. The special branch corresponding to $n=0$ requires an independent treatment, leading to logarithmic contributions to the metric functions. 
Together, these distinguished cases illustrate that the parameter $n$ does not simply control the magnitude of the corrections, but rather interpolates between qualitatively different gravitational regimes.
Having obtained the modified black hole geometry, we derived its thermodynamic properties through the Iyer--Wald formalism. A central result of this work is the  derivation of the entropy-area relation
$
S_W=S_{BH}+\lambda\,\Upsilon_n\,S_{ BH}^{\,n+1},
$
where the coefficient $\Upsilon_n$ depends only on the parameters of the underlying theory. This compact expression provides a unified description of the leading nonpolynomial curvature corrections to the BH entropy and naturally encompasses several distinguished limiting cases. 
{Although the functional dependence of $S_W$ coincides with that of the entropy-area relation previously obtained in the cosmological setting from the apparent-horizon thermodynamics~\cite{Dehyadegari:2026drv}, and  even though  we have functional agreement between $S_W$ and the apparent-horizon entropy, profound  differences arise for $n=-1$ and $n=1/2$. The differences in the corresponding coefficients  reflect and ambiguities in the thermodynamic construction at the apparent horizon—such (for example using the  Hawking temperature in dynamical cosmological scenario) as well as the fundamental difference distinction between Static/Killing horizons and dynamical/trapping horizons.}
In addition, the values $n=-1$ and $n=0$ exhibit an especially interesting behavior. For $n=-1$, the first-order Iyer--Wald entropy acquires an area-independent correction, while the corresponding metric perturbations remain non-trivial, indicating that the thermodynamic and dynamical sectors possess different limiting behaviors within the perturbative framework adopted here. Moreover, the neighborhood of this distinguished point naturally generates a logarithmic contribution as the leading correction to the constant entropy shift, revealing that constant, logarithmic, and power-law entropy corrections arise as different realizations of the same underlying entropy-area relation. By contrast, the $n=0$ branch simply rescales the BH entropy without modifying its linear dependence on the horizon area.

Finally, the present analysis demonstrates that this class of $f(\mathcal{R},\mathcal{K})$ theories remains tractable, at least perturbatively, despite the presence of nonpolynomial curvature interactions. The existence of closed perturbative black hole solutions together with a closed-form expression for the Iyer--Wald entropy provides a solid framework for further investigations, including the computation of the Hawking temperature and heat capacity, the analysis of black hole stability and quasinormal modes, the construction of rotating solutions, and the search for possible observational signatures associated with nonpolynomial curvature effects.}

\section*{Acknowledgements}
This work is supported by SECIHTI grants 257919, 258982 and by CIIC 027-2026. {\bf I.D.S.} is supported by the program {\it``Estancias Postdoctorales por México''}.  {\bf M. S.} is supported by the SECIHTI program  {\it ``Estancias sab\'aticas vinculadas a la consolidaci\'on de grupos de investigaci\'on''}. {\bf W. Y.} thanks the SECIHTI Grant CBF2023-2024-2923 {\it``Implications of the Generalized Uncertainty Principle (GUP) in Quantum Cosmology, Gravitation, and its Connection with Non-extensive Entropies''}, and is supported by SECIHTI/{\it``Estancias Postdoctorales por México''}. {\bf J.C.L-D} is supported by the grant UAZ-2024-39113. All authors are supported by the program {\it``Sistema Nacional de Investigadoras e Investigadores''}.


\bibliographystyle{apsrev4-1}
\bibliography{referencesp}
\end{document}